\documentclass{aa501}
\usepackage{graphicx}
\begin{document}
   \title{Thermoluminescence of Simulated Interstellar Matter
             after Gamma-ray Irradiation}

  \subtitle{Forsterite, Enstatite and Carbonates}


   \author{K. Koike\inst{1}
          \and
          M. Nakagawa\inst{1}
          \and
          C. Koike\inst{2}
          \and
          M. Okada\inst{3}
          \and
          H. Chihara\inst{4,2}
          }

   \offprints{K. Koike}

   \institute{Faculty of Education, Kagawa University,
             Takamatsu 760-8522, Japan\\
             \email{koike@ed.kagawa-u.ac.jp}
         \and
             Kyoto Pharmaceutical University,
             Kyoto 607-8412, Japan\
         \and 
             Research Reactor Institute, Kyoto University,
             Kumatori 590-0499, Japan\
         \and
             Department of Earth and Space Science,
             Osaka University, 
             Toyonaka, Osaka 560-0043, Japan\
             }


\abstract{
Interstellar matter is known to be strongly irradiated by radiation and 
several types of cosmic ray particles. Simulated interstellar matter, 
such as forsterite 
$\rm Mg_{2}SiO_{4}$, enstatite $\rm MgSiO_{3}$ and magnesite $\rm MgCO_{3}$
has been irradiated with the $\rm ^{60}Co$ gamma-rays in liquid nitrogen, 
and also 
irradiated with fast neutrons at 10 K and 70 K by making use of the
low-temperature irradiation facility of Kyoto University Reactor (KUR-LTL. 
Maximum fast neutron dose is
$10^{17}n_f{\rm /cm^{2}}$). 
After irradiation, samples are stored in liquid nitrogen for several months
to allow the decay of induced radioactivity.  We measured the luminescence 
spectra 
of the gamma ray irradiated samples during warming to 370K using a 
spectrophotometer. For the forsterite and magnesite, the spectra exhibit a 
rather intense peak at about 645 -- 655 nm and 660 nm respectively, whereas 
luminescence scarcely appeared in olivine sample. The spectra 
of forsterite is very similar to the ERE of the Red Rectangle.
   }
\maketitle
%

\section{Introduction}

Interstellar matter will be irradiated by electromagnetic energy and 
several cosmic ray particles, such as gamma rays, neutrons, 
protons and heavy-ions etc. 
Irradiation will cause some changes in these materials, especially 
to their optical properties. 
Investigation of this problem is expected to advance our understanding
of interstellar and circumstellar matter. 
Especially, it is well known that 
extremely large fluxes of neutrons and gamma-rays have been emitted 
during super-nova explosions and during
the so called Hayashi-phase in the early stage of protostellar systems. 
Moreover, interplanetary-dust is often irradiated by gamma rays and 
fast neutrons during the periods of flare activity on the sun, which 
is always repeated. 
Interstellar and circumstellar space is typically at extremely low temperature 
and is always irradiated 
over cosmological time-scale.

Though little is known about irradiation environment outside of our solar system,
it is natural to suppose that there are the regions with sufficient strength of 
irradiation to cause thermoluminescences, for example, a region of not so extremely 
far from a super nova explosion or 
not so far from a source of gamma ray emission, and/or long time irradiation 
at extremely low temperature.
The effect of this radiation will accumulate in the low temperature environment. 
It will be observed provided that the condition to release the accumulated 
energy is realized in circumstellar space. 
   \begin{figure*}
   \includegraphics[width=18cm]{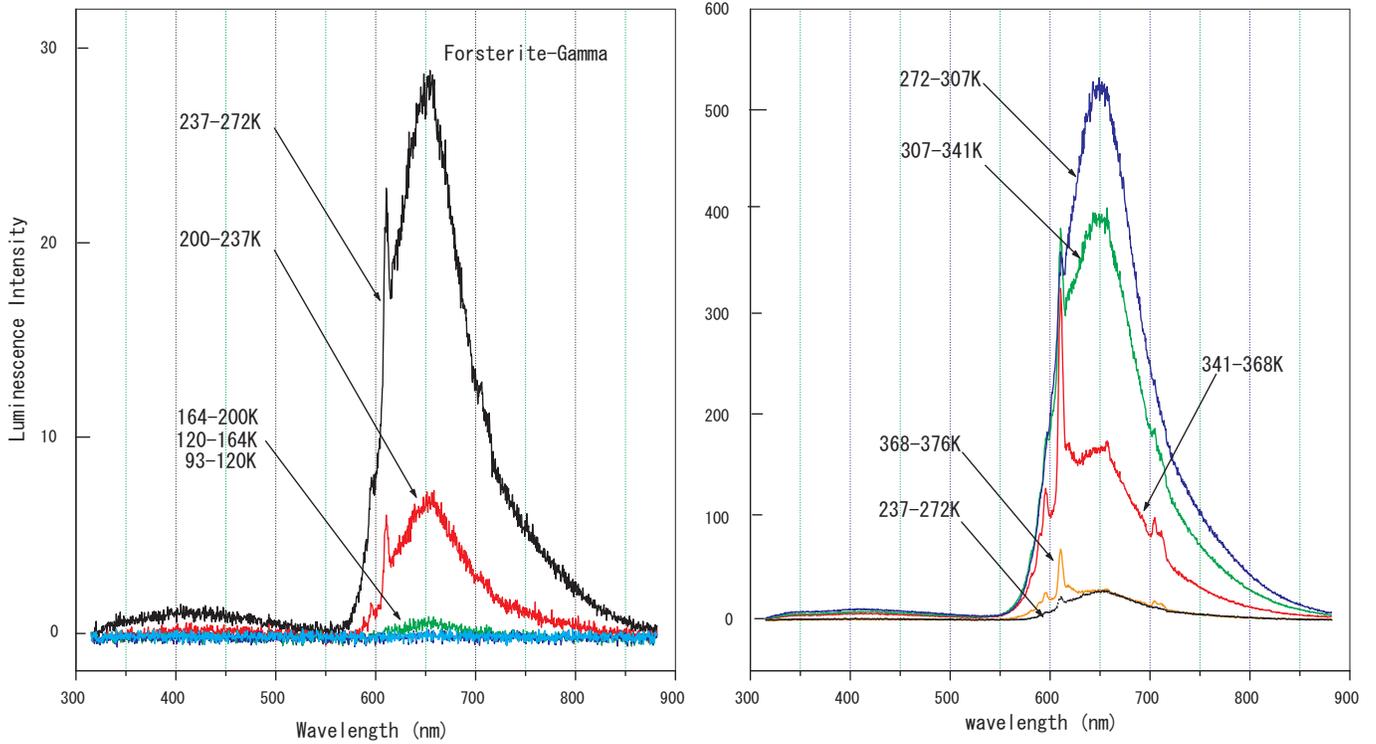}
   \caption{Thermoluminescence spectra of forsterite $(\rm Mg_2SiO_4)$ single 
         crystals. The left figure corresponds to 93-272K 
         and the right figure 272-376K.}
 \label{Forste_all}
    \end{figure*}
%
%

   \begin{figure}
   \centering
 \includegraphics[width=8.75cm]{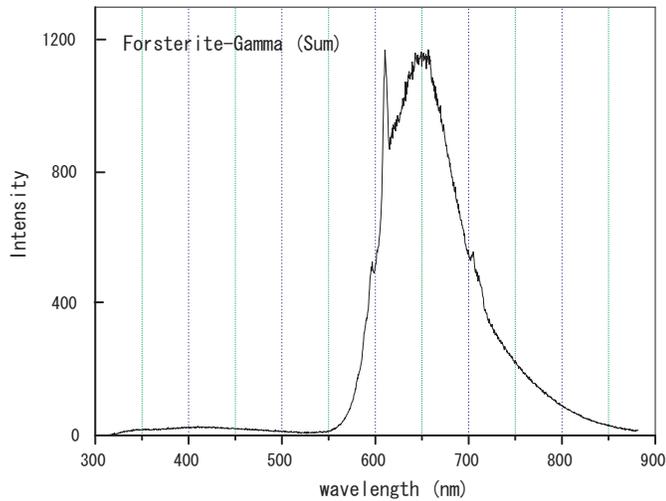}
  \caption{The sum of total thermoluminescences spectra of forsterite in Fig.1
  The peak of the thermoluminescence of 
$\rm Mg_{2}SiO_{4}$ is about 645-655 nm.}
 \label{Forste_sum}
    \end{figure}
%
   \begin{figure}
   \centering
 \includegraphics[width=8.75cm]{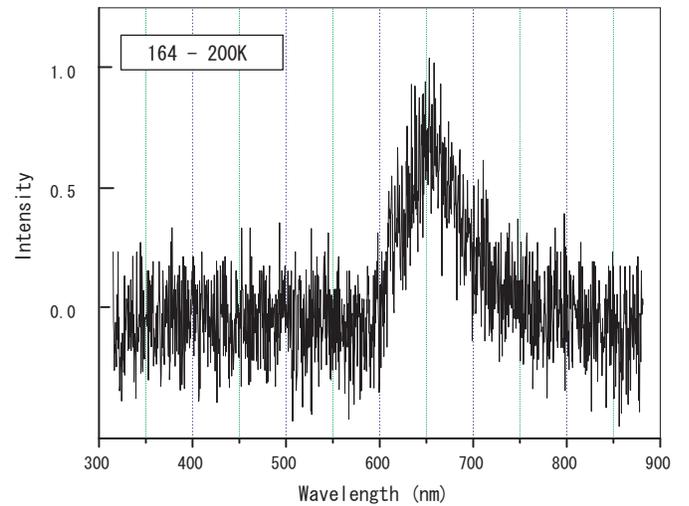}
 \caption{ The earliest (low temperature) thermoluminescences spectra 
of forsterite at 164-200K.
  The peak is also at about 645-655 nm.}
 \label{Forste016G-Low}
    \end{figure}
   The effects of irradiation on simple single crystals such as 
$\rm SiO_2, CaCO_3$ and $\rm CaF_2$ have been investigated by
several authors (Nakagawa et al. \cite{nakagawa1}, Nakagawa et al. 
\cite{nakagawa2}) from the viewpoint of solid state physics or material 
science. However, irradiation of materials of interstellar matter
has not been studied. It may then be worthwhile to 
investigate the effects of irradiation on simulated interstellar matter 
such as forsterite, enstatite and carbonates.  
Forsterite and enstatite have been found by 
many ISO observations in both young and evolved stars and in our own solar 
system (Waters et al.\cite{waters}, Malfait et al. \cite{malfait}, Wooden 
et al.\cite{wooden}). Carbonates such as dolomite ($\rm CaMg(CO_{3})_{2}$), 
breunnerite ($\rm Mg(Fe,Mn)(CO_{3})_{2}$), calcite ($\rm CaCO_3$), and 
$\rm Mg$, $\rm Ca$-bearing siderite ($\rm FeCO_3$) were found in CI 
chondrite (Endress, Zinner and Bischoff \cite{endress}). Especially,
it should be noted that the broad emission feature responsible for
extended red emission(ERE) 
appears at about the 500 -- 900 nm region in many reflection nebulaes,
and the Red Rectangle nebula shows sharp emission features over a broad 
band (Witt and Boroson \cite{witt}). In the Red Rectangle nebulae,
both PAH-features and crystalline silicates (forsterite and enstatite)
were observed (Waters et al., 1998).
The effect of irradiation on the optical properties of the 
simulated interstellar materials such as forsterite and enstatite 
by gamma rays and  neutrons  at low temperature are very interesting. 
%

   \begin{figure*}
   \includegraphics[bb=0 20 280 220,width=17.5cm]{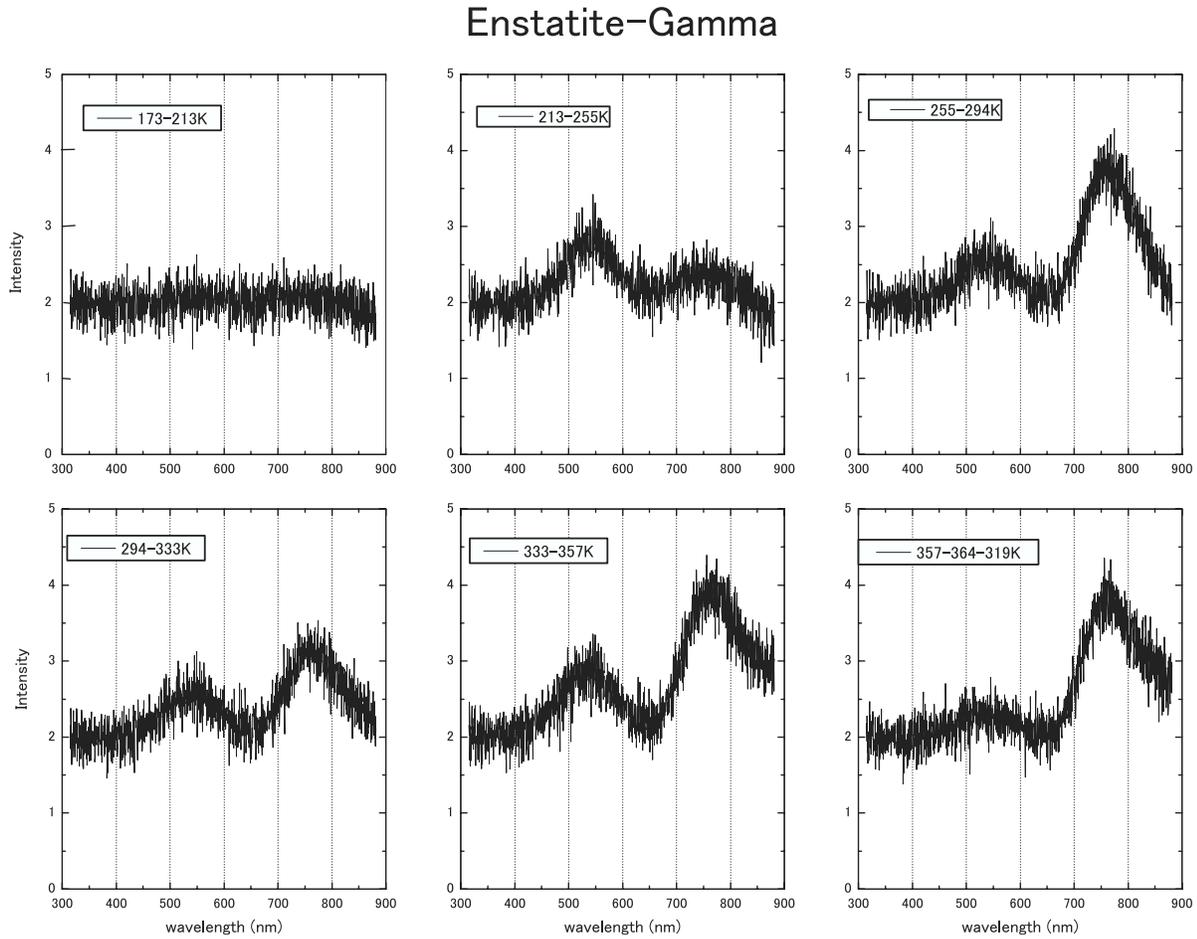}
   \caption{Thermoluminescence spectra of enstatite $(\rm MgSiO_3)$ single 
         crystals. (The down right-side figure shows the luminescence when 
         the temperature rose from 357K to  maximum 364K,
         and come down till 319K)  }
 \label{Enstatite}
 \end{figure*}
   \begin{figure*}
   \sidecaption
   \includegraphics[width=13.5cm]{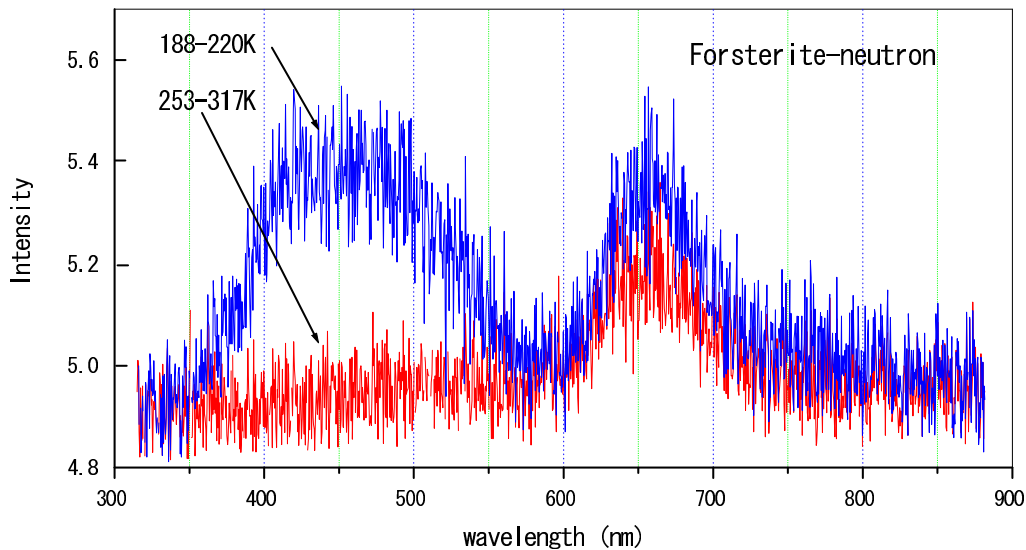}
 \caption{~The thermoluminescences spectra 
of forsterite after neutron irradiated at 10K.
 \vspace{5.15cm}\noindent}
 \label{Forste_n}
    \end{figure*}
%
%
   \begin{figure*}
   \sidecaption
 \includegraphics[bb= 0 0 218 140,width=12.25cm]{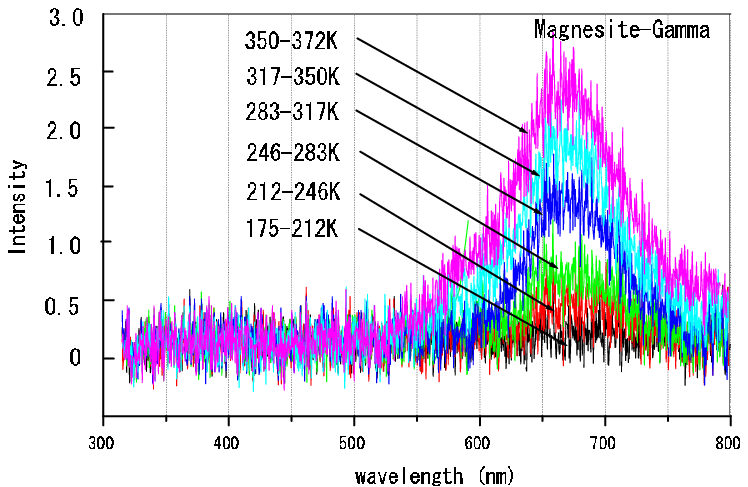}
   \caption{ Thermoluminescence spectra of magnesite $(\rm MgCO_3)$.
   The peak is at about 660nm.                                      
 \vspace{6.75cm}\noindent}
 \label{MgCO3_all}
    \end{figure*}

\section{Features of Thermoluminescence spectrum}

  Bulk samples such as forsterite, $\rm Mg_2SiO_4$, natural olivine from Egypt, 
orthoenstatite $\rm MgSiO_3$ and natural magnesite $\rm MgCO_3$ from the 
Democratic People's Rep. of Korea were 
irradiated with gamma-rays to a dose of about $10.4 \times 10^4$ Gy $\rm(J/Kg)$
in liquid nitrogen 
using the $\rm ^{60}Co$ gamma-ray irradiation facility of Kyoto University 
Reactor. The 
gamma-rays of $\rm ^{60}Co$ have two peaks at 1.1 MeV and 1.3 MeV. 
Our samples of forsterite and enstatite were synthesized by Takei and 
Kobayashi (1974), and
Tachibana (2000) using the CZ (Czochralski) and Flux method, respectively   
with high accuracy. The bulk of the irradiated forsterite is triangular-shaped
(about $6\times8\times15$ mm size, about 1 mm thickness) and weights about 138 mg. 
The irradiated enstatite consists of several 
small fragments (about 1 - 2 mm size) and the total weight is about 114 mg.

We have measured the thermoluminescence 
spectra of these samples using a spectrophotometer
(including a CCD camera, Princeton Instruments, Inc.). 
The sample is put on the thermally-isolated plate, 
which has previously been cooled to liquid nitrogen temperature. 
The luminescence emission during warming is introduced to the CCD measuring 
system using an optically transparent fiber.
The color of forsterite changed to dark-white and -gray 
(rather dark red-violet) after gamma-ray irradiation. 
The time of warming the samples from liquid nitrogen temperature to room 
temperature (333 K) is about 15 min. It should be noted that this time length
is sufficient to observe the thermoluminescence, that is, photon is emitted at 
the thermal equilibrium state.

 Fig.\ref{Forste_all} shows the luminescence spectra 
of forsterite $(\rm Mg_2SiO_4)$ after Gamma-ray irradiation. 
It shows a rather intense broad peak at about 645--655 nm 
(and a weak broad peak at 400--440 nm),
 and a sharp peak at 610 nm. 
The sharp peak at 610 nm appears as the sample warms from 200K 
to room temperature. Other weak peaks at 590, 595, and 705 nm
appear distinctly at above 200 K. 
The broad peak at 645-655 nm becomes suddenly strongest at 272 -- 307 K  
, and the weak peaks fade into background.
Above 307 K, the luminescence becomes weak gradually and another peaks at 595,
605, 705, 710 nm become prominent.
The previous preliminary data of another sample of irradiated forsterite, its weight 
about 244 mg, have shown almost the same luminescence spectra, but it seems that
the earliest luminescence starts at somewhat low temperature.
Fig.\ref{Forste_sum} shows the total strength of luminescence in 
Fig.\ref{Forste_all}.  
A magnification of the emitted thermoluminescences spectrum of the most 
low-temperature at the range about 164-200K in  Fig.\ref{Forste_all} 
is shown in Fig.\ref{Forste016G-Low}. 
   \begin{figure}
 \includegraphics[width=7.55cm]{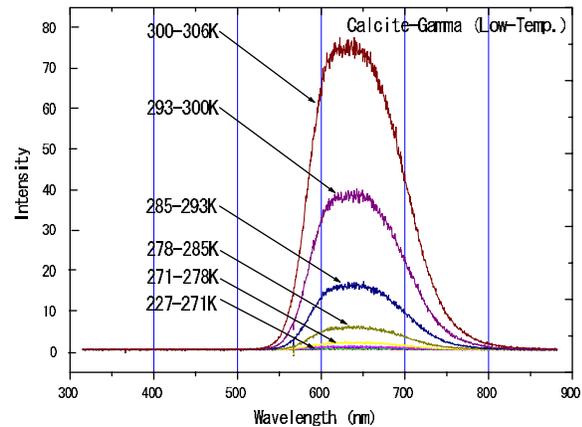}
  \caption{Typical thermoluminescence spectrum of calcite ($\rm CaCO_3$)
  at 227-300K.
 The peak of $\rm CaCO_3$ is at about 640 nm.}
 \label{CaCO3_L}
    \end{figure}
%
   \begin{figure}
 \includegraphics[width=7.55cm]{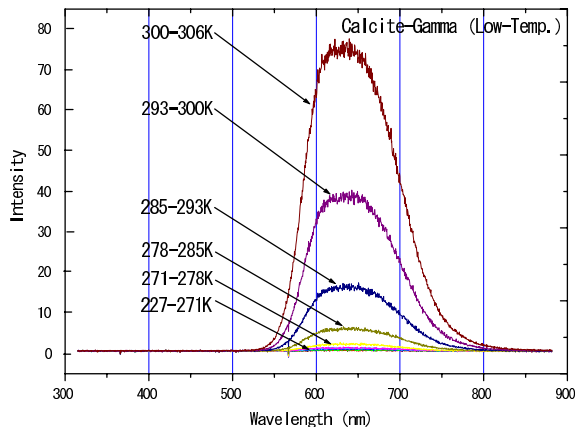}
  \caption{Typical thermoluminescence spectrum of calcite ($\rm CaCO_3$)
  at 300-337K. The peak shifts toward to 600nm.}
 \label{CaCO3_H}
    \end{figure}

In contrast, the luminescence is scarcely visible in natural olivine
(from Egypt, $\rm Fo_{90} $). 
As for enstatite, we have detect 
the luminescence spectra, which is shown in Fig.\ref{Enstatite}. 
Two broad peaks appear at about 545 and 760 nm, 
and these peaks are very weak compared with the spectra of forsterite.
The weight of enstatite used was 114 mg, while that of forsterite 
was 138 mg as mentioned before.

 Fig.\ref{MgCO3_all} shows the measurement of luminescence spectra of magnesite
$\rm MgCO_3$ after Gamma-ray irradiation.  
The spectrum of $\rm MgCO_3$ shows a very broad peak at about 660nm, and 
this peak becomes strong at 350 -- 375 K. 
The irradiated magnesite is a bulk sample 
(about $5\times5\times2$ mm size) and its weight is 93 mg.
It should be emphasized that
the thermoluminescence of $\rm MgCO_3$ also has sufficient strength to be 
visible easily. 
The color of magnesite changed to a rather white-gray with 
pink-yellow tendency after gamma-ray irradiation. 

We have also measured the thermo-dependence of luminescence of calcite.
The present luminescence of calcite is shown in Fig.\ref{CaCO3_L} 
and  Fig.\ref{CaCO3_H}.
 The intensity becomes stronger as temperature increased from 77 K, and the 
peak position is 620-640 nm at 300 K. Above 300 K, the intensity became further 
stronger, and at 327 - 337 K the intensity became strongest, the shape of 
luminescence changed and the peak position shifts to around 605 nm. Higher 
temperature than 337 K, the intensity became weak.

\section{Related problems and the other measurements}
We also investigated the effect of irradiation on both a bulk sample 
and particle samples  embedded in polyethylene 
using neutrons to a dose of  $8 \times 10^{16} n_f{\rm /cm^{2}}$  
at 10 K and 70 K using the 
low-temperature irradiation facility of Kyoto University Reactor
(KUR-LTL: Low Temperature Loop)(Okada et al., 2001). 
After neutron irradiations of 75 hour, 
samples are stored in liquid nitrogen for several months to wait a 
decay of radioactivity.

We measured the absorption coefficient of a neutron-irradiated 
forsterite and enstatite particles using a FTIR spectrometer(Nicolet, 
Nexus 670) 
over the wavelength range from 30 -- 100 $ \mu m$. The samples were 
irradiated 
at 70 K using a neutrons in September, 2000. The gross 
feature of spectrum show no apparent difference from the 
spectrum (Koike C. et al. \cite{koikeC1}, \cite{koikeC2}) taken 
before irradiation. One of the reasons for this result may be that
the fast neutrons strongly collide with hydrogen atoms in the 
polyethylene and only weakly collide with sample particles
dispersed in polyethylene.

 We have also measured the thermoluminescence spectrum of neutron-irradiated 
bulk sample of forsterite, which is shown in  Fig.~\ref{Forste_n}.
At low-temperature, two peaks appeared at about 450 and 650 nm, but above 
253K one peak at 450nm became weak.
The peak at 650nm is almost the same as Gamma-ray irradiated case, however, 
the strength of luminescence is considerably weak. This reason is not known yet.

For future irradiation experiments using a stronger beam of neutrons in the
center of the reactor, we are interested in an accurate measurement of the 
impurity component in our forsterite sample, because impurity component with
long a half period brings serious difficulties in measurements of irradiated 
samples. Previously, Takei and Kobayashi
( \cite{takei}) reported that the spectrographic analysis shows that the 
forsterite sample is pure, 
but using neutron activation analysis, an infinitesimal quantity of Ir at
about 16--18 wt ppm had been found. This level of Ir impurity is below the 
detection limit of the spectrographic analysis.
We have also measured our sample using radio activation analysis, 
and confirmed that our 
sample is almost pure; that is, other elements except for $\rm Mg, Si, O$ 
and an infinitesimal quantity of Ir are not detected.

\section{Similarity to ERE of Red Rectangle Spectrum}

   \begin{figure}
   \centering
 \includegraphics[width=7.5cm]{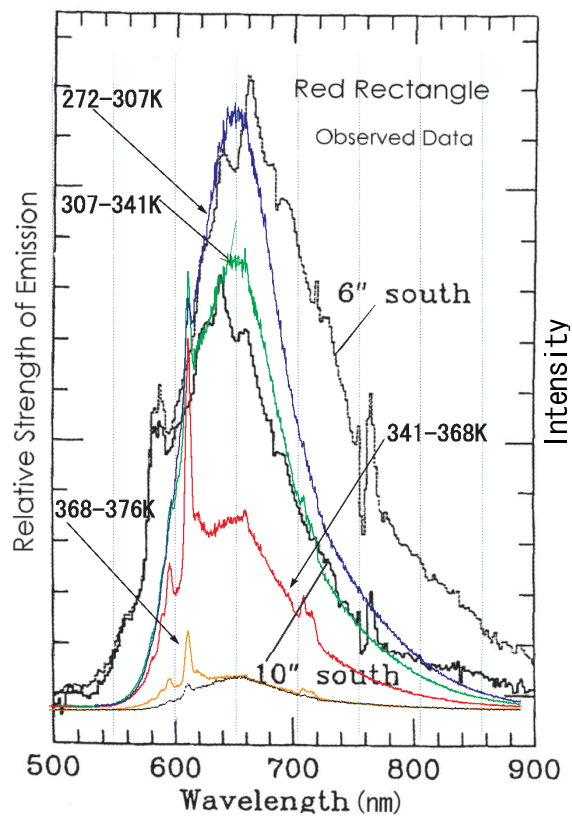}
 \caption{Comparison with Observation Data of Red Rectangle Spectrum
  (From Witt A.N. and Boroson T.A. 1990) and Forsterite Spectrum (the right-side of
  Fig.1) }
 \label{Red-Rect2}
    \end{figure}

It is remarkable fact that the observed spectrum of ERE (Extended Red Emission) 
of the Red 
Rectangle (Witt and Boroson \cite{witt}) given in Fig.~\ref{Red-Rect2} shows 
characteristic feature. 
As for carriers of ERE, possible size effect of crystalline silicon 
nanoparticles has been discussed (Ledoux G. et al. \cite{Ledoux}). 
It has been shown that the photoluminescence spectra of these nanoparticles 
can explain the gross structure of ERE spectra, such as the peak position
and the full width at half maximum of ERE spectra etc.
However, evidence of existence of crystalline silicon has not yet been observed. 
(Li and Draine \cite{Li1},\cite{Li2})

 We have interested that the luminescence of forsterite is very similar to ERE of Red Rectangle (peak at 652-684 nm, half width 70-90 nm) among many objects. Generally, ERE from interstellar dust consists of a broad, featureless emission band peaking at 610-820 nm, and width 60-100 nm. Among many objects, Red Rectangle show strong emission of one order stronger and exhibits the characteristic feature. Furthermore, crystalline silicates such as forsterite and PAH dust had been founded in Red Rectangle by ISO observations. The present irradiated forsterite also shows very strong luminescence (peak at 640-660 nm) at about 280-308 K and half width of about 100 nm,
and characteristic peaks at 590, 595, and 705nm.

It should be also remembered that forsterite and enstatite have been found by many 
ISO observations in many oxygen rich young and evolved stars.
However, ERE is observed in many carbon rich stars. It is important to note that
Red Rectangle is very interesting object with both carbon and oxygen rich characteristics,
that is, both forsterite and PAH were found.
  It should be emphasized that our thermoluminescence spectrum of forsterite at 
about 645--655 nm  and at 590 nm is very similar to the ERE of the Red Rectangle.

\section{Discussion}

For the irradiation circumstance, little is known around ERE object. However,
it should be noted that interstellar and circumstellar space is typically at 
extremely low temperature and is always irradiated by electromagnetic radiation 
and by cosmic ray particles over cosmological time-scale. 
Furthermore, it is well known that extremely large fluxes of neutrons and 
gamma-rays have been emitted during super-nova explosions.
The effect of this radiation will accumulate in the low temperature environment. 
It will only be observed provided that the condition to release the accumulated 
energy is realized in circumstellar space. This may be occur when irradiated dust 
move to a warmer domain in interstellar or circumstellar environment.

The accumulated energy in irradiated matter is released by thermoluminescence
when the matter is warmed. The rate of this release is dominated by Boltzman 
factor, and in extremely low temperature its rate is practically infinite.
In connection with this problem, it is interesting fact that the thermoluminescence 
spectrum of particles in Tyrrhenian Sea exhibits the typical peak at the layer of 
year 1054, 1006, 1181 etc., corresponding to the year of super-nova explosions
(Castagnoli G. C. et al., \cite{Castagnoli}). 
In their estimate, the energy emitted during super-nova explosions of Crab nebula 
 at year 1054 is about $10^{47} \rm{J}$ and the estimated flux of energy at the point
of distance of the radius of the Galaxy is about $10^5 \rm{J} /m^2$. 

Though most energies of explosions are carried by neutrinos, a certain rate will be 
carried by X and gamma rays.
It will be possible in 
extremely low-temperature environment such as interstellar or circumstellar space,
the effects of irradiation are "frozen" and almost stable in extremely long time, 
and various irradiation effects, such as often observed super-nova explosions or
always irradiated cosmic rays, are accumulated.

The irradiation on matter generally cause lattice defects of crystal,
and a certain kind of this effect is observed in thermoluminescence. It is known
in many experiences that the irradiation effect on solid can be erased by 
annealing it to  several hundred degrees. In other word, the effects of irradiation 
remains in a certain kind of form below this temperature and it may be observed still 
in effects different from the thermoluminescence.

\section{Summary}
In this paper, we report the data for gamma-ray irradiation
of some simulated interstellar materials. Forsterite $(\rm Mg_2SiO_4)$,
enstatite  $(\rm MgSiO_3)$
and magnesite $(\rm MgCO_3)$ exhibit interesting thermoluminescences spectra. 

  We have emphasized that the thermoluminescence of forsterite 
appears at low temperature at about 160-200K. 
In particular, it should be emphasized that the spectrum of forsterite at about 
645--655 nm  and at 590 nm is very similar to the ERE of the Red Rectangle.
(Witt and Boroson \cite{witt})
It should be noted that possible size effect of crystalline silicon 
nanoparticles can explain the gross structure of ERE spectra
(Ledoux G. et al. \cite{Ledoux}).
However, this model can explain only the gross structure of spectrum of ERE,
and seems not to explain the characteristic features of 590nm. 
 The most interesting fact of size effect
is the peaks of gross structure of luminescence shifts depending on the size 
of nanoparticles. 
It is interesting
problem to examine possible existence of similar effect in forsterite 
nanoparticles.

Finally, the irradiation on matter is expected to cause various effects on matter in 
addition to thermoluminescence effects.
Investigating these effects in the context of astrophysics is further 
problem.


\begin{acknowledgements}
  The authors would like to express their sincere thanks to Prof.
  Tsuchiyama A., 
  who provided superior samples of simulated interstellar matter 
  such as forsterite and enstatite.
  We would like also to acknowledge members of Prof. Atobe's group of Naruto 
  University of Education for using their equipment including the sample holder 
  with thermo controller.
  
 This work was supported by the KUR projects (12062, 13p1-6).
 Part of this work was supported by Grant-in-Aid of Japanese Ministry of 
 Education, Science, and Culture (12440054). 
\end{acknowledgements}


\begin{thebibliography}{}

\bibitem[1982]{Castagnoli} Castagnoli G.C., Bonino G., and Miono S. 1982,
        IL Nuovo Cimento, 5C, 488

\bibitem[1996]{endress} Endress M., Zinner E. \& Bischoff A. 1996,
        Nature, 379, 701

\bibitem[2000]{koikeC1} Koike C. Tsuchiyama A., Shibai H. et al. 2000,
        A\&A, 363, 1115

\bibitem[2000]{koikeC2} Koike C., Chihara H., Tsuchiyama A. et al. 2000,
        Proc. 33rd ISAS Lunar and Planet. Symp. 33, 95

\bibitem[2001]{Ledoux} Ledoux G., Guillois O., Huisken F. et al. 2001,
        A\&A, 377, 707
        
\bibitem[2001]{Li1} Li A. and Draine B.T. 2001,
        ApJ, 550, L213
        
\bibitem[2001]{Li2} Li A. and Draine B.T. 2002,
        ApJ, 564, 803
        
\bibitem[1998]{malfait} Malfait K., Waelkens C., Waters L.B.F.M. et al. 1998,
        A\&A, 332, L25
        
\bibitem[1988]{nakagawa1} Nakagawa M., Fukunaga M., Okada M. et al. 1988,
       Journal of Luminescence 40\&41, 345

\bibitem[1999]{nakagawa2} Nakagawa M., Koike K., Okada M. e al., 1999,
        KURRI Prog. Rep. 1999, 77

\bibitem[2001]{okada} Okada M., Kanazawa S., Nozaki T. et al., 2001, 
        Nuclear Instruments and Methods in Physics Research A 463, 213

\bibitem[2000]{tachibana} Tachibana S., 2000, Ph. D. Thesis, Osaka 
       University

\bibitem[1974]{takei} Takei H. and Kobayashi T. 1974,
      J. Crystal Growth, 23, 121
      
\bibitem[1998]{waters} Waters L.B.F.M., Beintem D.A., 
      Zijlstra A.A. et al. 1998,
      A\&A, 331, L61

\bibitem[1998]{waters} Waters L.B.F.M., waelkens C., Van Winckel H. et al.,
     1998, Nature, 391, 868


\bibitem[1990]{witt} Witt A.N. and Boroson T.A. 1990,
        ApJ, 517, 1034

\bibitem[1999]{wooden} Wooden D.H., Harker D.E., Woodward C.E. et al. 1999,
        ApJ, 355, 182
        
\end{thebibliography}
\end{document}